\documentclass[runningheads]{llncs}
\usepackage[T1]{fontenc}
\usepackage{graphicx}
\usepackage{caption}
\usepackage{subcaption}
\usepackage{booktabs}
\usepackage{adjustbox}
\usepackage{svg}
\usepackage{amsmath}
\usepackage{amssymb}
\usepackage{mathtools}
\usepackage{amsthm}
\usepackage{amsfonts}

\newcommand{\beginsupplement}{
  \setcounter{table}{0}  
  \renewcommand{\thetable}{S\arabic{table}} 
  \setcounter{figure}{0} 
  \renewcommand{\thefigure}{S\arabic{figure}}
}

\usepackage{hyperref}
\usepackage{color}

\usepackage[capitalize,noabbrev]{cleveref}

\usepackage[symbol]{footmisc}

\begin{document}

\title{Multimodal LLMs for health grounded in individual-specific data}
\titlerunning{Multimodal LLMs for health}

\author{
Anastasiya Belyaeva \inst{1}\thanks{Equal contribution; authors listed alphabetically.} % \orcidID{0000-1111-2222-3333}
\and
Justin Cosentino\inst{1}$^*$ % \orcidID{0000-1111-2222-3333}, \footnotemark[1] is broken in this template.
\and
Farhad Hormozdiari \inst{2} % \orcidID{0000-1111-2222-3333}
\and
Krish Eswaran \inst{1} % \orcidID{0000-1111-2222-3333}
\and
Shravya Shetty \inst{1} % \orcidID{0000-1111-2222-3333}
\and
Greg Corrado \inst{1} % \orcidID{0000-1111-2222-3333}
\and
Andrew Carroll
 \inst{1} % \orcidID{0000-1111-2222-3333}
\and
\\ Cory Y. McLean \inst{2}\thanks{Equal supervision.} % \orcidID{0000-1111-2222-3333}
\and
Nicholas A. Furlotte \inst{1}$^\dagger$ % \orcidID{0000-1111-2222-3333}
}
\authorrunning{Belyaeva et al.}
\institute{
Google Research, San Francisco CA 94105, USA 
\and
Google Research, Cambridge MA 02142, USA \\
\email{nickfurlotte@google.com} \\
% \email{\{belyaeva,jtcosentino,fhormoz,cym,nickfurlotte\}@google.com}
}

\maketitle
\begin{abstract}
Foundation large language models (LLMs) have shown an impressive ability to solve tasks across a wide range of fields including health. 
To effectively solve personalized health tasks, LLMs need the ability to ingest a diversity of data modalities that are relevant to an individual's health status.
In this paper, we take a step towards creating multimodal LLMs for health that are grounded in individual-specific data by developing a framework (HeLM:  Health Large Language Model for Multimodal Understanding) that enables LLMs to use high-dimensional clinical modalities to estimate underlying disease risk. HeLM encodes complex data modalities by learning an encoder that maps them into the LLM's token embedding space and for simple modalities like tabular data by serializing the data into text.
Using data from the UK Biobank, we show that HeLM can effectively use demographic and clinical features in addition to high-dimensional time-series data to estimate disease risk. For example, HeLM achieves an AUROC of 0.75 for asthma prediction when combining tabular and spirogram data modalities compared with 0.49 when only using tabular data. Overall, we find that HeLM outperforms or performs at parity with classical machine learning approaches across a selection of eight binary traits. Furthermore, we investigate the downstream uses of this model such as its generalizability to out-of-distribution traits and its ability to power conversations around individual health and wellness.

\end{abstract}

\section{Introduction}
\label{introduction}

Foundation large language models (LLMs) have been shown to solve a range of natural language processing (NLP) tasks without having been explicitly trained to do so \cite{brown2020language,wei2022emergent}. As a result, researchers are adapting LLMs to solve a variety of non-traditional NLP problems across domains. A recent perspective \cite{moor2023foundation} outlined a variety of health-related use cases that could benefit from foundation LLMs that have not only generalist medical knowledge but that are also infused with individual-specific information such as lab values (e.g., cholesterol and triglycerides), imaging, time-series data, health tracker metrics (e.g., daily step count and heart rate), genome sequences, genetic risk scores, and other omics data modalities. These use cases range from AI clinician assistants to AI-powered early warning systems to user-facing health and wellness chatbots.

While the potential applications for foundation LLMs in health are wide-ranging, at the core of each there is a fundamental need for the model to ingest complex multimodal individual-specific data and use it to gain an understanding of the individual's underlying health risks. The model can then condition responses to queries on the derived risk profile of an individual. 
Though there has been promising recent work in developing generalist medical LLMs \cite{yang2022large,singhal2022large,singhal2022medpalm2,steinberg2021language}, the problem of using multimodal individual-specific information as context for health-related tasks remains understudied.
More broadly, this capability represents one aspect of the general movement towards personalization of LLMs \cite{kirk2023personalisation,salemi2023lamp}, which encompasses not only the technical challenges of data integration, but also the complex ethical questions around how the model can and should be used.

\begin{figure*}[t]
% \vskip 0.2in
\centering
%\includesvg[width=1\textwidth]{figures/overview2.svg}
\includegraphics[width=1\textwidth]{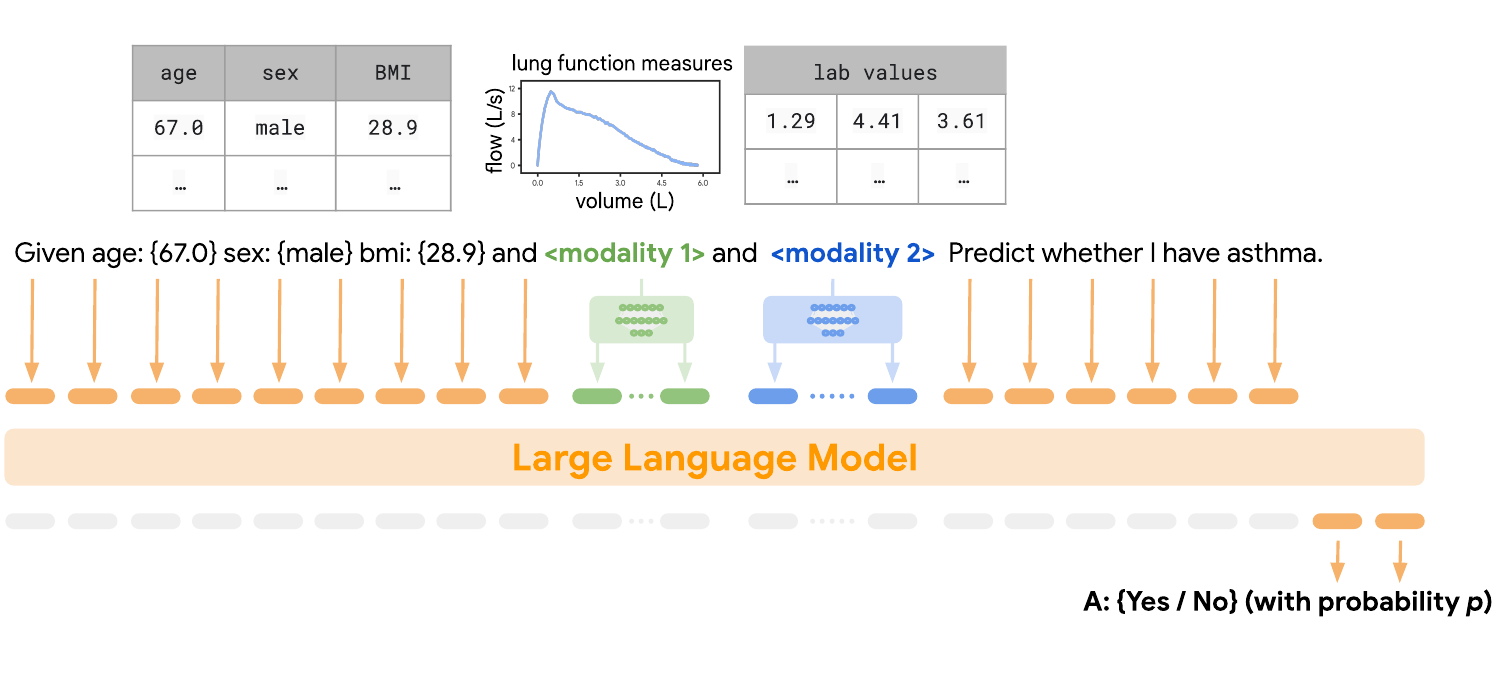}
% \vspace*{-8mm}
\caption{\textbf{Overview of HeLM, a multimodal LLM for health}. Text features (orange) are tokenized 
and embedded into the token embedding space via a standard embedding matrix. Non-text modalities such as clinical data (blue) or high-dimensional lung function measures (green) are encoded into the same token embedding space via modality-specific encoders. The LLM is tuned to quantify disease risk given complex multimodal inputs.}
\label{fig:overview}
% \vskip -0.2in
\end{figure*}

Providing relevant health information to an LLM could be as simple as including important disease risk factors, such as lab values, in the prompt by representing these factors as text \cite{hegselmann2023tabllm,dinh2022lift}. Furthermore, in-context learning techniques such as few-shot prompting \cite{brown2020language} can be employed by giving the model examples that help it to connect risk factors to underlying disease. However, this solution isn't likely to work for the many complex data modalities found in the health space \cite{acosta2022multimodal}. For example, it isn't clear how to incorporate images, time-series data, or even rich structured tabular data into the LLM prompt. Furthermore, health factors may not be clearly captured by a single modality but rather may be best predicted by simultaneously incorporating features drawn from multiple modalities.

Recently, a variety of methods have been introduced to extend LLMs to the multimodal setting. The majority of these methods have focused on images and text \cite{radford2021CLIP,li2023blip2,alayrac2022flamingo,jia2021scaling,yu2022coca,openai2023gpt4} with some models adding the ability to incorporate diverse modalities, such as the movements of robots, video, and audio \cite{driess2023palm,girdhar2023imagebind,lu2022unified}. However, these nascent methods have not yet been applied to the health domain. On the other hand, there are a variety of classical machine learning methods for integrating multiple data modalities that are routinely applied in the health domain \cite{acosta2022multimodal}. Examples include logistic regression classifiers that take multiple modalities as input, various fusion models \cite{kline2022multimodal}, autoencoder-based models \cite{yangbelyaeva2021multi}, and cross-supervision models \cite{zhou2022multimodalradiology}. 
However, these traditional approaches lack several potential advantages when compared to LLM-based methods. 

First, foundation LLMs may have encoded extensive prior knowledge about health related traits. For example, an LLM likely understands that hypertension is the same as high blood pressure and may even know something about what number ranges correspond to---normal or high. This prior knowledge can be useful when dealing with heterogeneous data or with data that only has fuzzy labels. Secondly, LLMs can incorporate additional prior knowledge through prompt engineering, whereas in traditional ML methods including priors can be cumbersome. Thirdly, LLMs may have a high degree of flexibility in working with missing data. Whereas traditional methods require imputation of missing values or dropping samples, LLMs can be prompted to ignore missing values or they can be omitted completely without architectural changes. Finally, many foundation LLMs are conversational by design, so they can more naturally be used for applications such as those mentioned previously: user-facing health and wellness chatbots.

In this paper, we take a step towards multimodal LLMs for health that are grounded in individual-specific data by developing a framework called HeLM that enables LLMs to use health-related features to estimate underlying disease risk. The idea is that an LLM with a representation of background risk, can use this as context in answering health-related queries. 
We formulate a disease classification task and then use the LLM, Flan-PaLMChilla 62b \cite{chung2022scaling}, to score potential outcomes. We compare the LLM score with classic supervised machine learning methods such as logistic regression and gradient-boosted decision trees by evaluating their ability to distinguish between individuals with and without the disease.

Using data from the UK Biobank \cite{Bycroft2018}, we evaluate different ways of prompting and encoding data and passing it to the LLM to classify disease status.
First, we serialize health-related features into text similar to previously proposed approaches \cite{hegselmann2023tabllm,dinh2022lift}. We evaluate zero-shot, few-shot, and a parameter-efficient soft-prompt tuning approach \cite{Lester2021}.
Generally, we find that the LLM performs better than random in the zero-shot and few-shot cases, and has comparable and often equivalent performance to standard approaches such as logistic regression and a gradient-boosted decision trees (implemented in XGBoost \cite{chen2016xgboost}) after soft-prompt tuning. Additionally, for some diseases (e.g., diabetes), the zero-shot and few-shot approaches perform surprisingly well when compared with logistic regression and XGBoost, giving evidence to the model's use of prior knowledge related to disease risk.
However, we observe that performance degrades with an increase in the number of input features indicating that the model does not always fully capture signal from serialized text.

To better capture signal from quantitative data, we propose HeLM, a multimodal approach to disease risk estimation based on the PaLM-E framework \cite{driess2023palm}. In HeLM, non-text modalities are mapped via encoders (trained over input examples) into the same token embedding space as text, where a separate encoder is learned for each modality (see \Cref{fig:overview}). We show that HeLM matches the performance of classical machine learning methods (logistic regression and XGBoost) and, in some cases, outperforms both the logistic regression and XGBoost models. We show results for encoding tabular data and spirogram curves, a time series representation of lung function used for respiratory disease diagnosis \cite{Vestbo2013,cosentino2023copd}. Finally, we explore how HeLM that has been tuned to quantify disease risk can be used in downstream conversational tasks. Here, we find that conversational ability degrades in the tuned LLM, which is consistent with what others have reported \cite{wang2022preserving}. We discuss future directions such as fine-tuning schemes that may mitigate degradation.

\section{Methods}
\label{section:methods}

\subsection{LLMs with Tabular Data}
An individual's health status is often represented by a set of values stored in tabular format. We explore whether serializing tabular data into natural language and passing the resulting text to the LLM (Flan-PaLMChilla 62b \cite{chung2022scaling}) is sufficient to achieve strong predictive performance. We construct serialized inputs by mapping table column names and corresponding values into JSON-like format, which is combined with the base prompt. For example, for diabetes prediction, we formulate the following sentence using the data: \textit{“Predict if a patient has the condition or not. bmi: \{28.9\}. age: \{67.0\}. sex: \{male\}. diabetes: \{”}. We then compute the log-likelihood of the sentence being completed with \textit{“yes\}”} or \textit{“no\}”}. This log-likelihood serves as a risk score and can be evaluated using metrics such as AUROC and AUPRC to assess discriminatory power.

We evaluate the LLM’s performance in the zero-shot, few-shot and soft-prompt tuning settings. In the zero-shot setting, the serialized text input is given directly to the model. In this case, the LLM heavily leverages prior knowledge. In the few-shot scenario, we prefix 10 examples randomly sampled from the training dataset to the model’s prompt. For soft-prompt tuning, following \cite{Lester2021}, we learn a soft prompt to condition the frozen language model to perform well on the diseases of interest.

Briefly, soft-prompt tuning is a parameter efficient tuning technique that's commonly used to steer the frozen LLM to perform well on a downstream task given labeled data. Instead of fine-tuning the weights of the LLM or forming a hard prompt out of tokens as in the few-shot scenario, soft-prompt tuning learns a matrix $P \subset \mathbb{R}^{p \times k}$ in the token embedding space, where $p$ is the length of the prompt and $k$ is the size of the language embedding space. The soft-prompt $P$ is then concatenated with the embedded tokens and passed to the decoder. We train the soft-prompt using pairs of examples $(X, Y)$ and backpropagation to update the soft-prompt with the goal of maximizing the probability of $Y$. We train for 20,000 steps and use 1000 training examples.

\subsection{Multimodal LLMs for health: HeLM}
To enable the LLM to reason over complex high-dimensional inputs, we embed non-text data modalities, including time-series data like spirograms and tabular data, into the same latent space as the language tokens (see \cref{fig:overview}). We use separate encoders for each non-text data modality, where each encoder learns a mapping to the token embedding space. This approach is based on the PaLM-E framework \cite{driess2023palm}. All of the inputs are then processed together by a pre-trained LLM. More precisely, LLMs typically tokenize the words in a sentence and map the resulting tokens $w_i$ into a language embedding space $\mathcal{X} \subset \mathbb{R}^k$ via a large embedding matrix $x_i = \gamma(w_i)$ where $\gamma:\mathcal{W} \to \mathcal{X}$.  In HeLM, non-text modalities are mapped to a sequence of vectors in $\mathcal{X}$ via encoders. For example, for mapping the spirogram time-series we trained $\phi_s:\mathcal{S} \to \mathcal{X}^{q_s}$ encoder and for mapping tabular data we trained $\phi_t:\mathcal{T} \to \mathcal{X}^{q_t}$ encoder, where $q_s$ and $q_t$ correspond to the number of vectors in $\mathcal{X}$ space or in other words how many ``multimodal'' tokens each modality is mapped to. We set $q_s = 6$ and $q_t =6$ for all experiments, however in general these can be treated as tunable hyper-parameters. In summary, each vector $x_i$ is formed from either the token embedder $\gamma$ or a modality-specific encoder:

\begin{equation*}
x_i = 
\begin{cases} 
\gamma(w_i) & \text{if } i \text{ is a text token} \\
\phi_s(s_i) & \text{if } i \text{ corresponds to the spirogram modality} \\
\phi_t(t_i) & \text{if } i \text{ corresponds to the tabular modality}
\end{cases}
\end{equation*}

Non-text and text modalities can be injected in any order. We train the modality-specific encoders keeping the pre-trained LLM weights frozen. This is similar to soft-prompt tuning but conditioned on data from a specific modality. For experiments considering multiple diseases we train a single HeLM model on a mixture of all diseases as opposed to one model per disease.

\subsection{UK Biobank Dataset Preparation}
\label{section:methods:ukbiobank}

We obtain clinical features, spirometry, and disease labels from the UK Biobank (UKB) \cite{Bycroft2018}. Similar to previous work, we focus on the European genetic ancestry subpopulation \cite{alipanahi2021large}. Limiting to European ancestry within the UK Biobank is a standard heuristic for reducing phenotypic heterogeneity, due to the correlation between population structure and phenotypic variation \cite{diaz2019umap}. Differences in phenotypes across ancestries are multifactorial---socio-economic, cultural, etc.---but are often highly correlated with genetic background and thus population structure. Therefore, selecting study individuals based on a single genetic background is a convenient way to reduce heterogeneity in underlying disease risk, at the expense of creating bias in the dataset and subsequent analyses that utilize this data.

We defined binary phenotype labels using medical records comprising ICD-9 hospital inpatient (HESIN) billing codes, ICD-10 primary care and general practitioner (GP) read codes, and self-report data. Diseases include asthma, diabetes, hypertension, stroke, myocardial infarction, major depression, migraine, all cause mortality, cataracts, gastroesophageal reflux disease (GERD), hay fever eczema (atopic eczema), osteoarthritis, and pneumonia. The following clinical features, lab values, and self-reported statuses sourced from questionnaires were used as model inputs: age, sex, body mass index (BMI), high-density lipoprotein (HDL) cholesterol, low-density lipoprotein (LDL) cholesterol, total cholesterol, triglycerides, diastolic blood pressure, smoking status, snoring, insomnia, daytime napping, average nightly sleeping, and chronotype. See \Cref{appendix:table:phenotypes} for details describing feature definitions.

Spirometry data was prepared following the preprocessing procedures outlined in \cite{cosentino2023copd}. In short, we used raw volumetric flow curves containing exhalation volume sampled at 10-ms intervals to liters and then computed the corresponding flow curve by approximating the first derivative with respect to time by taking a finite difference. The volume-time and flow-time curves were then normalized to length 1,000 and combined to generate a one-dimensional flow-volume spirogram. Following well-accepted spirometry quality control standards, we filter the dataset to individuals with at least one acceptable blow from the first visit \cite{sakornsakolpat2019copd,shrine2019lung,cosentino2023copd}.

We randomly partitioned patients with valid data entries for all phenotype labels and clinical features into distinct training and validation datasets.

\section{Experimental Results}
\label{section:results}
We present LLM risk prediction performance across eight disease classification tasks under varied experimental settings. We begin by assessing the effectiveness of zero-shot and few-shot prompting as well as soft-prompt tuning on a pre-trained foundation model using health data serialized into text. We then demonstrate that directly mapping quantitative features into the LLM's latent token space significantly improves model performance. Finally, using asthma as a case study, we show that this mapping procedure generalizes to high-dimensional lung function data.

\subsection{Quantifying disease risk using zero-shot, few-shot, and soft-prompt tuning}
\label{section:results:prompting}
We first establish a baseline for LLM disease risk prediction using zero-shot, few-shot, and soft-prompt tuning with a frozen Flan-PaLMChilla 62b model \cite{chung2022scaling}.
We define classification tasks using a diverse set of binary phenotype targets (see \cref{section:methods:ukbiobank}). For each task and prompting method, we evaluate a ``baseline'' set of model inputs consisting of age, sex, and BMI as well as an ``expanded'' set that includes eleven additional clinical and wellness features: HDL cholesterol, LDL cholesterol, total cholesterol, total triglycerides, diastolic blood pressure, smoking status, average sleep duration, insomnia, snoring, daytime napping, and chronotype (i.e., whether a patient is a morning or evening person). These predictors were chosen based on prior knowledge that they should be informative about the selected targets. 

We score a validation dataset ($n=3,000$) using the methodology outlined in \Cref{section:methods}. For each validation sample, we generate a disease risk score by computing the log probability of a positive disease label. Similarly, to obtain the logistic regression and XGBoost scores, we compute the probability of a positive disease label given the respective model fit on a separate training set ($n=10,000$).

\Cref{table:llm_logreg_perf} shows performance for each prompting method and input set compared to the baseline models. We observe that at least one prompting technique is competitive with the baselines for most tasks. In some cases (e.g., hypertension and diabetes), zero-shot and few-shot models perform surprisingly well despite seeing little to no training data compared to the baselines, an observation also made by \cite{hegselmann2023tabllm}. This suggests that the LLM uses prior knowledge about the relationships between age, sex, BMI and disease likelihood. 

\begin{table*}[ht!]
\caption{\textbf{Comparison of AUC and AUPRC between LLM-based classifiers and classical machine learning approaches on the validation set.} The models with ``baseline'' input features use age, sex and BMI as model features, while the models with the ``expanded'' set also include 11 additional clinical and wellness features. The mean AUC/AUPRC and the corresponding 95\% confidence intervals were calculated across 1,000 bootstrapping iterations. Bold cells denote the best models for a given phenotype and input feature set, where statistical significance is determined via paired bootstrapping. Logistic regression and XGBoost models were trained on 10,000 samples, few-shot on 10 samples and soft-prompt tuning on 1,000 samples.}
\label{table:llm_logreg_perf}
% \vskip 0.15in
\begin{center}
\begin{small}
\begin{adjustbox}{max width=\textwidth}
\begin{tabular}{llllll}
\toprule
\textbf{Phenotype} & \textbf{Model} & \textbf{Baseline AUC} & \textbf{Expanded AUC} & \textbf{Baseline AUPRC} & \textbf{Expanded AUPRC} \\
\midrule
All Cause Mortality & Zero-shot & 0.61 (0.57--0.64) & 0.61 (0.57--0.65) & 0.10 (0.08--0.12) & 0.11 (0.09--0.14) \\
(prevalence = 6.83\%) & Few-shot & 0.65 (0.61--0.69) & 0.65 (0.61--0.69) & 0.12 (0.10--0.15) & \textbf{0.12 (0.09--0.15)} \\
 & Soft-prompt & \textbf{0.69 (0.66--0.73)} & \textbf{0.69 (0.65--0.72)} & 0.13 (0.10--0.16) & \textbf{0.14 (0.11--0.18)} \\
 & LogReg & \textbf{0.71 (0.67--0.74)} & \textbf{0.68 (0.64--0.72)} & \textbf{0.15 (0.12--0.19)} & \textbf{0.16 (0.13--0.21)} \\
 & XGBoost & 0.67 (0.62--0.70) & \textbf{0.67 (0.62--0.70)} & \textbf{0.13 (0.10--0.17)} & \textbf{0.13 (0.10--0.17)} \\
\midrule
Diabetes & Zero-shot & 0.70 (0.67--0.73) & 0.61 (0.57--0.65) & 0.18 (0.14--0.23) & 0.12 (0.10--0.15) \\
(prevalence = 7.60\%) & Few-shot & 0.72 (0.69--0.76) & 0.67 (0.64--0.70) & \textbf{0.19 (0.15--0.24)} & 0.14 (0.11--0.17) \\
 & Soft-prompt & 0.72 (0.69--0.76) & 0.68 (0.64--0.72) & \textbf{0.23 (0.18--0.28)} & 0.17 (0.14--0.22) \\
 & LogReg & \textbf{0.74 (0.70--0.77)} & \textbf{0.75 (0.72--0.79)} & \textbf{0.23 (0.18--0.28)} & \textbf{0.26 (0.21--0.32)} \\
 & XGBoost & \textbf{0.73 (0.69--0.76)} & 0.73 (0.69--0.76) & \textbf{0.22 (0.17--0.26)} & 0.22 (0.17--0.26) \\
\midrule
Hypertension & Zero-shot & 0.70 (0.68--0.72) & 0.68 (0.66--0.70) & 0.60 (0.56--0.62) & 0.57 (0.54--0.60) \\
(prevalence = 40.03\%) & Few-shot & 0.73 (0.71--0.75) & 0.72 (0.70--0.73) & 0.62 (0.59--0.64) & 0.59 (0.56--0.62) \\
 & Soft-prompt & 0.72 (0.70--0.74) & 0.72 (0.70--0.74) & 0.60 (0.57--0.63) & 0.59 (0.56--0.62) \\
 & LogReg & 0.74 (0.72--0.76) & 0.74 (0.72--0.76) & 0.63 (0.60--0.66) & \textbf{0.65 (0.62--0.68)} \\
 & XGBoost & \textbf{0.77 (0.75--0.78)} & \textbf{0.77 (0.75--0.78)} & \textbf{0.66 (0.63--0.69)} & \textbf{0.66 (0.63--0.69)} \\
\midrule
Major Depression & Zero-shot & 0.54 (0.51--0.57) & 0.58 (0.55--0.61) & 0.15 (0.13--0.17) & 0.17 (0.14--0.19) \\
(prevalence = 13.03\%) & Few-shot & 0.50 (0.47--0.53) & 0.55 (0.52--0.58) & 0.14 (0.12--0.15) & 0.15 (0.13--0.17) \\
 & Soft-prompt & \textbf{0.60 (0.57--0.63)} & 0.47 (0.44--0.50) & \textbf{0.20 (0.17--0.23)} & 0.12 (0.11--0.14) \\
 & LogReg & \textbf{0.60 (0.57--0.63)} & \textbf{0.62 (0.59--0.65)} & \textbf{0.19 (0.16--0.22)} & \textbf{0.19 (0.17--0.22)} \\
 & XGBoost & 0.54 (0.52--0.57) & 0.54 (0.52--0.57) & 0.16 (0.14--0.18) & 0.16 (0.14--0.18) \\
\midrule
Migraine & Zero-shot & 0.51 (0.45--0.56) & 0.49 (0.45--0.55) & 0.05 (0.03--0.07) & 0.04 (0.03--0.05) \\
(prevalence = 3.77\%) & Few-shot & 0.45 (0.40--0.50) & 0.47 (0.42--0.53) & 0.03 (0.03--0.04) & 0.03 (0.03--0.04) \\
 & Soft-prompt & \textbf{0.64 (0.59--0.69)} & 0.51 (0.46--0.56) & \textbf{0.07 (0.05--0.09)} & 0.04 (0.03--0.06) \\
 & LogReg & \textbf{0.62 (0.56--0.68)} & \textbf{0.62 (0.56--0.67)} & \textbf{0.07 (0.04--0.11)} & \textbf{0.06 (0.04--0.08)} \\
 & XGBoost & \textbf{0.60 (0.54--0.66)} & \textbf{0.60 (0.54--0.66)} & \textbf{0.06 (0.04--0.07)} & \textbf{0.06 (0.04--0.07)} \\
\midrule
Myocardial Infarction & Zero-shot & 0.61 (0.57--0.65) & 0.61 (0.56--0.65) & 0.08 (0.06--0.10) & 0.09 (0.07--0.12) \\
(prevalence = 5.93\%) & Few-shot & 0.67 (0.63--0.71) & 0.65 (0.61--0.68) & \textbf{0.11 (0.09--0.14)} & 0.10 (0.08--0.12) \\
 & Soft-prompt & \textbf{0.71 (0.67--0.74)} & \textbf{0.68 (0.64--0.71)} & \textbf{0.12 (0.10--0.15)} & 0.11 (0.09--0.14) \\
 & LogReg & \textbf{0.71 (0.67--0.74)} & \textbf{0.70 (0.65--0.74)} & \textbf{0.12 (0.09--0.15)} & \textbf{0.16 (0.12--0.21)} \\
 & XGBoost & \textbf{0.69 (0.65--0.73)} & \textbf{0.69 (0.65--0.73)} & \textbf{0.14 (0.11--0.19)} & \textbf{0.14 (0.11--0.19)} \\
\midrule
Stroke & Zero-shot & 0.61 (0.56--0.66) & 0.55 (0.49--0.60) & 0.06 (0.05--0.09) & 0.05 (0.04--0.06) \\
(prevalence = 4.00\%) & Few-shot & 0.65 (0.60--0.69) & 0.60 (0.56--0.65) & 0.07 (0.05--0.09) & 0.05 (0.04--0.07) \\
 & Soft-prompt & 0.63 (0.58--0.68) & 0.50 (0.45--0.55) & \textbf{0.09 (0.06--0.12)} & 0.05 (0.03--0.06) \\
 & LogReg & \textbf{0.69 (0.64--0.74)} & \textbf{0.73 (0.68--0.77)} & \textbf{0.10 (0.07--0.14)} & \textbf{0.14 (0.09--0.19)} \\
 & XGBoost & \textbf{0.68 (0.63--0.72)} & 0.68 (0.63--0.72) & \textbf{0.09 (0.06--0.12)} & 0.09 (0.06--0.12) \\
\bottomrule
\end{tabular}
\end{adjustbox}
\end{small}
\end{center}
\vskip -0.2in
\end{table*}

Focusing on the baseline feature set (age, sex, and BMI), we aimed to understand how the LLM derives scores. We estimated the importance of each input feature by regressing the features against the scores output by the model. Coefficients from the linear regression model are used to measure feature importance and concordance across methods. \Cref{appendix:fig:lr_compare_baseline} shows the result of this analysis for four traits. For diabetes, hypertension, and stroke, we see concordance between logistic regression, XGBoost and the LLM models in terms of direction and relative magnitude of effects. Additionally, we find a strong correlation between logistic regression and LLM scores (Spearman = 0.46--0.93 across prompting methods), while the correlation between LLM and XGBoost scores is weaker (Spearman = 0.39--0.65).
On the other hand, for migraine, we see little concordance in direction and relative magnitude of effects for zero-shot and few-shot LLMs, indicating that the LLM doesn't have sufficient prior knowledge to relate migraine with the input features. However, this is corrected in the soft-prompt case, where we see concordance and high Spearman correlation between soft-prompt tuned LLM and logistic regression (0.85). On the other hand, the soft-prompt tuned LLM has low Spearman correlation with XGBoost (0.31), which is a non-linear model. 
Given this, we hypothesize that the LLM is effectively scoring outcomes using what translates to a simple linear function of the input features and that this linear mapping is effectively learned via soft-prompting.

In the expanded feature set, we often find that the LLM with zero-shot prompting has poorer performance (e.g., on diabetes) when compared with the LLM with zero-shot prompting using the smaller feature set. 
This may indicate that the LLM has insufficient prior knowledge about the features contained in the expanded set.
However, unlike the baseline features, the model is not able to match baseline performance even when using soft-prompt tuning.
This indicates that for a complex set of input features, text serialization does not yield a representation that fully captures the available signal.
Together, these results motivate the use of a multimodal approach to directly embed quantitative data into the prompt.

\subsection{Encoding quantitative data using HeLM}
\label{section:results:helm_quantitative}

To assess the benefit of directly embedding quantitative data into the LLM's latent token space, we repeat the experiments from \Cref{section:results:prompting} but encode the ``extended'' inputs using the HeLM framework. We learn an encoder $\phi_t$, an MLP (two hidden layers of size 1024 and 4096, respectively), that takes as input quantitative features and maps them into the embedding space. We train this model over a mixture of all seven binary traits ($n=10,000$ for each trait).

\Cref{table:multimodal_llm_logreg_perf} shows performance metrics for HeLM compared with logistic regression and XGBoost. 
We see that by directly encoding the tabular data into token space, HeLM performs at parity with the best baseline methods and for all cause mortality, hypertension and myocardial infarction HeLM outperforms the best baseline. In comparing the scoring methods, we again find that HeLM scores are highly correlated with logistic regression scores (Spearman = 0.70--0.87; \cref{appendix:fig:corr_all_cause_mortality}-\cref{appendix:fig:corr_stroke}). In \cref{appendix:fig:lr_compare}, we repeat the feature importance analysis, this time comparing HeLM with logistic regression and XGBoost. Similar to \cref{section:results:prompting}, we see strong concordance between feature weights, particularly for features that have the most importance. Taken together, these results are consistent with our previous hypothesis that the LLM scores outcomes in a way that is highly consistent with logistic regression and that the LLM is arriving at these scores by similarly linearly weighting input features.

In the case of hypertension, we see that HeLM significantly outperforms logistic regression and XGBoost, while XGBoost appears to have a slight advantage over logistic regression.
In addition, the HeLM score is more correlated with the XGBoost score when compared with logistic regression (Spearman 0.87 vs. 0.76), which is not the case for any of the other traits.
Although, in no way conclusive, this may indicate that by mapping the tabular data into the token embedding space, the LLM takes advantage of non-linear relationships between input features and hypertension.
This phenomenon may account for some of the cases where HeLM outperforms the baseline methods.

We hypothesized that HeLM has an advantage over logistic regression due to being trained over a mixture of traits and thus benefiting from transfer learning across traits. To evaluate this, we selected three traits (hypertension, all cause mortality and myocardial infarction) where HeLM performed better than logistic regression in terms of AUROC. For each, we trained a single task HeLM (a separate model for each trait) and compared with the mixture. Overall, we see that the mixture trained HeLM does not have an advantage over the single task model (\cref{appendix:table:mixture_vs_single_task}). Training on a mixture of diseases is still advantageous since it yields a single model that can be used for a variety of diseases as opposed to separate models for each disease (as is the case for logistic regression and XGBoost).

\begin{table}
\caption{\textbf{Comparison of AUC and AUPRC between a HeLM, logistic regression and XGBoost models on the validation set.} The binary phenotypes are predicted from the 14 ``expanded set'' input features from Table 1. The features are both encoded as text as in Table 1 and also included as a secondary quantitative data modality.}
\label{table:multimodal_llm_logreg_perf}
% \vskip 0.15in
\begin{center}
\begin{small}
\begin{adjustbox}{max width=\columnwidth}
\begin{tabular}{llll}
\toprule
\textbf{Phenotype} & \textbf{Model} & \textbf{AUC} & \textbf{AUPRC} \\
\midrule
All Cause & HeLM & \textbf{0.71 (0.68--0.75)} & \textbf{0.18 (0.14--0.23)} \\
Mortality & LogReg & 0.68 (0.64--0.72) & \textbf{0.16 (0.13--0.21)} \\
 & XGBoost & 0.67 (0.62--0.70) & 0.13 (0.10--0.17) \\
\midrule
Diabetes & HeLM & \textbf{0.77 (0.74--0.80)} & \textbf{0.27 (0.22--0.33)} \\
 & LogReg & \textbf{0.75 (0.72--0.79)} & \textbf{0.26 (0.21--0.32)} \\
 & XGBoost & 0.73 (0.69--0.76) & 0.22 (0.17--0.26) \\
\midrule
Hypertension & HeLM & \textbf{0.79 (0.77--0.81)} & \textbf{0.69 (0.66--0.72)} \\
 & LogReg & 0.74 (0.72--0.76) & 0.65 (0.62--0.68) \\
 & XGBoost & 0.77 (0.75--0.78) & 0.66 (0.63--0.69) \\
\midrule
Major & HeLM & \textbf{0.63 (0.60--0.65)} & \textbf{0.19 (0.17--0.22)} \\
Depression & LogReg & \textbf{0.62 (0.59--0.65)} & \textbf{0.19 (0.17--0.22)} \\
 & XGBoost & 0.54 (0.52--0.57) & 0.16 (0.14--0.18) \\
\midrule
Migraine & HeLM & \textbf{0.61 (0.55--0.66)} & \textbf{0.06 (0.04--0.07)} \\
 & LogReg & \textbf{0.62 (0.56--0.67)} & \textbf{0.06 (0.04--0.08)} \\
 & XGBoost & \textbf{0.60 (0.54--0.66)} & \textbf{0.06 (0.04--0.07)} \\
\midrule
Myocardial & HeLM & \textbf{0.73 (0.69--0.77)} & \textbf{0.15 (0.12--0.19)} \\
Infarction & LogReg & 0.70 (0.65--0.74) & \textbf{0.16 (0.12--0.21)} \\
 & XGBoost & 0.69 (0.65--0.73) & \textbf{0.14 (0.11--0.19)} \\
\midrule
Stroke & HeLM & \textbf{0.73 (0.68--0.77)} & \textbf{0.14 (0.09--0.20)} \\
 & LogReg & \textbf{0.73 (0.68--0.77)} & \textbf{0.14 (0.09--0.19)} \\
 & XGBoost & 0.68 (0.63--0.72) & 0.09 (0.06--0.12) \\
\bottomrule
\end{tabular}
\end{adjustbox}
\end{small}
\end{center}
\vskip -0.2in
\end{table}

\subsection{Estimating asthma risk using multiple modalities}
\label{section:results:helm_spiro}

Using HeLM to leverage tabular data clearly showed that this is a promising direction for quantifying disease risk. Next, we evaluate whether HeLM can incorporate more complex data modalities such as a spirogram, a time-series curve which measures the amount of air an individual can breath in and out of their lungs. Spirometry is commonly used to assess pulmonary function and the presence of respiratory diseases. Thus, we focus on the task of quantifying asthma risk.

To this end, we trained a HeLM model with three modalities as input: tabular data (14 ``expanded set'' input features) as well as spirometry, along with tabular data serialized to text. We do not include a textual description of the spirogram since it's unclear how to summarize this data in text.
We encode spirometry data into the token embedding space via a one-dimensional variant of the ResNet18 architecture \cite{he2016deep,he2019bag}, followed by an MLP (two hidden layers of size 1024 and 4096, respectively). We use a pre-trained model from \cite{cosentino2023copd} for the ResNet18 part of the encoder and only update the weights of the MLP. The ResNet18 model was trained to predict asthma and COPD. We take as input the 128-dimensional embedding, corresponding to the penultimate layer. Similar to \cref{section:results:helm_quantitative}, we use an MLP to encode tabular data into the token embedding space.

The model is instructed to predict asthma using supervised labels (``yes'' or ``no'') as targets in training. We trained on $n = 16,724$ samples from the UK Biobank, which were obtained by subsampling the dataset to achieve a one-to-one case-control class distribution to ensure a balanced representation of categories. As a direct baseline, we also trained a model that linearly combined the 128-dimensional embedding from the ResNet18 model and tabular data, which we term ResNet18 1D (tabular data $+$ spirogram).

In order to assess whether the multimodal HeLM model is leveraging the additional spirogram modality for asthma prediction, we trained several models on tabular data only, without the spirogram. Following \cref{section:results:helm_quantitative} we trained a logistic regression model, XGBoost, an LLM with soft-prompt tuning and HeLM using tabular data only.

\begin{figure*}[ht]
% \vskip 0.2in
\begin{center}
\centerline{\includegraphics[width=\linewidth]{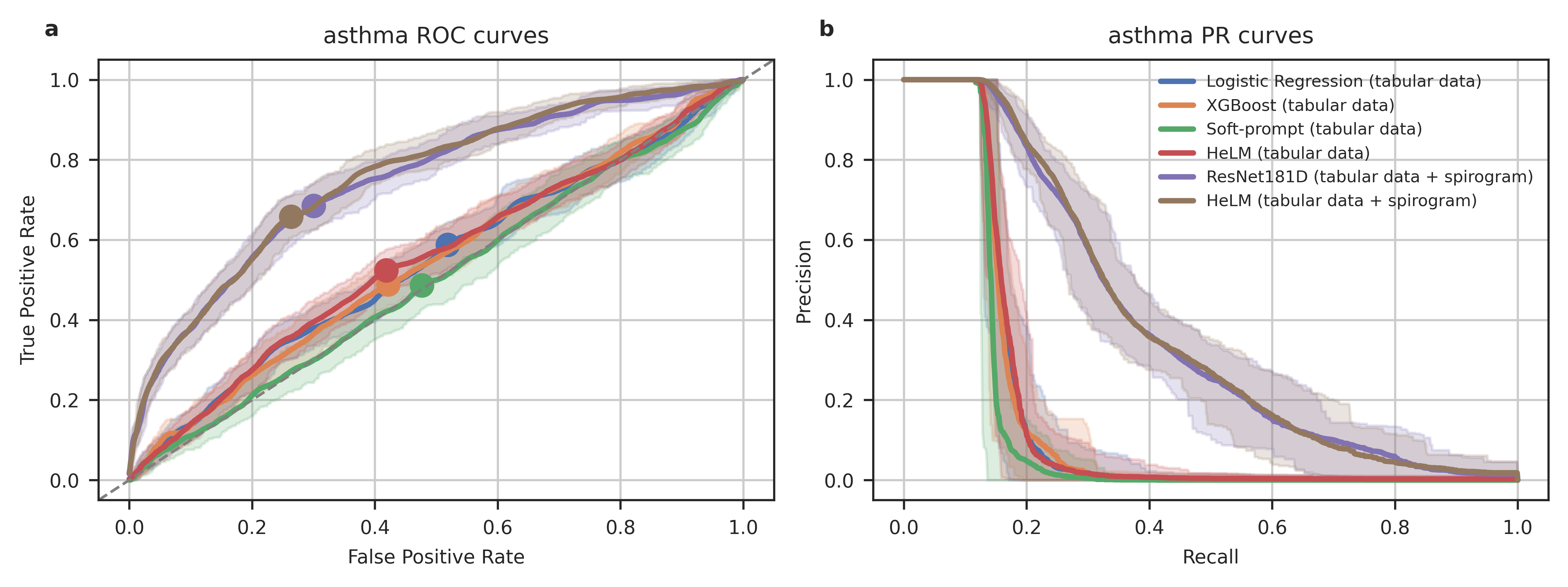}}
\caption{\textbf{Inclusion of additional input modalities improves HeLM asthma detection.} (a) ROC and (b) Precision-recall (PR) for asthma phenotype prediction using tabular data only versus tabular data and spirometry. We compare HeLM, ResNet18 1D, logistic regression, XGBoost, and a soft-prompt tuned LLM trained on either a single modality (tabular) or two modalities (tabular and spirometry).}
\label{fig:helm_spiro}
\end{center}
\vskip -0.3in
\end{figure*}

\Cref{fig:helm_spiro} shows the model performance on a held-out set of individuals with valid spirograms ($n = 2,289$). We observe that the HeLM model trained on two non-text modalities (tabular data and spirogram) is successfully leveraging the additional spirogram modality to boost the performance on asthma prediction. For example, the AUROC (AUPRC) increases from $0.49 \pm 0.02$ ($0.14 \pm 0.01$) for the LLM trained on tabular data only via soft-prompt tuning to $0.75 \pm 0.01$ ($0.38 \pm 0.02$) for HeLM trained on tabular data and a spirogram. The comparison with soft-prompt tuned LLM on tabular data is particularly important since it is the only other method that can also generate natural language text and thus provide recommendations, answer questions, and summarize given an individual's health context. We also observe that HeLM trained on tabular and spirogram data performs on par with the linear combination of the ResNet18 1D model on spirogram and tabular data.

\subsection{Using HeLM for out-of-distribution traits}
\label{section:results:unseen_traits}
Deep learning methods have shown significant performance degradation when applied to out-of-distribution (OOD) data \cite{recht2018cifar}. We investigated HeLM OOD performance by applying it to a set of traits not in the training set. Considering cataract, gastroesophageal reflux disease (GERD), hay fever eczema (atopic eczema), osteoarthritis, and pneumonia, we compared HeLM with trait-specific logistic regression models (\cref{table:unseen_traits}). We observed that HeLM (not trained on the trait) performs on par with logistic regression (trained on the trait) in four out of five of the tested traits.
Taken at face-value this implies that HeLM is leveraging prior information about the impact of risk factors on disease learned from related traits in the training data.
However, to truly understand how the model arrives at the OOD scores, will take additional research.

\begin{table}[t]
\caption{\textbf{Comparison of AUC and AUPRC between zero-shot, few-shot LLM-based classifiers, HeLM and a logistic regression model on the validation set in OOD setting.} While logistic regression models were trained on the data for the trait, HeLM was not trained for the trait. The binary phenotypes are predicted from the 14 ``expanded set'' input features from Table 1. The features are both encoded as text as in Table 1 and also included as a secondary quantitative data modality. ``prev'' denotes prevalence.}
\label{table:unseen_traits}
% \vskip 0.15in
\begin{center}
\begin{small}
\begin{adjustbox}{max width=\columnwidth}
\begin{tabular}{llll}
\toprule
\textbf{Phenotype} & \textbf{Model} & \textbf{AUC} & \textbf{AUPRC} \\
\midrule
Cataract & Zero-shot & 0.58 (0.55--0.61) & 0.17 (0.15--0.20) \\
(prev=13.51\%) & Few-shot & 0.63 (0.60--0.65) & 0.18 (0.16--0.20) \\
 & HeLM & \textbf{0.71 (0.69--0.73)} & 0.24 (0.21--0.27) \\
 & LogReg & \textbf{0.72 (0.69--0.74)} & \textbf{0.27 (0.23--0.31)} \\
\midrule
GERD & Zero-shot & 0.58 (0.55--0.60) & 0.22 (0.19--0.25) \\
(prev=17.33\%) & Few-shot & 0.57 (0.54--0.60) & 0.21 (0.18--0.23) \\
 & HeLM & \textbf{0.61 (0.59--0.64)} & \textbf{0.25 (0.22--0.28)} \\
 & LogReg & \textbf{0.61 (0.59--0.64)} & \textbf{0.26 (0.23--0.29)} \\
\midrule
Hay Fever & Zero-shot & 0.48 (0.46--0.51) & 0.23 (0.21--0.24) \\
Eczema & Few-shot & 0.46 (0.44--0.49) & 0.23 (0.21--0.25) \\
(prev=23.84\%) & HeLM & 0.47 (0.44--0.49) & 0.22 (0.20--0.24) \\
 & LogReg & \textbf{0.57 (0.55--0.60)} & \textbf{0.28 (0.26--0.31)} \\
\midrule
Osteoarthritis & Zero-shot & 0.60 (0.58--0.62) & 0.36 (0.33--0.39) \\
(prev=28.13\%) & Few-shot & 0.63 (0.61--0.65) & 0.38 (0.35--0.40) \\
 & HeLM & \textbf{0.65 (0.63--0.67)} & 0.41 (0.38--0.44) \\
 & LogReg & \textbf{0.67 (0.65--0.69)} & \textbf{0.44 (0.40--0.47)} \\
\midrule
Pneumonia & Zero-shot & 0.56 (0.52--0.60) & 0.08 (0.06--0.10) \\
(prev=6.50\%) & Few-shot & 0.59 (0.55--0.63) & 0.08 (0.07--0.10) \\
 & HeLM & \textbf{0.68 (0.65--0.72)} & \textbf{0.16 (0.12--0.21)} \\
 & LogReg & \textbf{0.67 (0.63--0.71)} & \textbf{0.16 (0.12--0.20)} \\
\bottomrule
\end{tabular}
\end{adjustbox}
\end{small}
\end{center}
\vskip -0.2in
\end{table}

\subsection{Natural language generation}
\label{section:results:generation}

Finally, we assess whether an LLM can incorporate multimodal data to tailor conversational tasks. Approximating a two-stage approach in which a model is trained to quantify risk and then trained to use these risk estimates in conversational tasks, we first use HeLM to compute asthma risk from a spirogram embedding and encode the predicted risk as text into the PaLM-2 model \cite{google2023palm2} for exercise recommendations: \textit{``I have a p\% chance of having asthma. What are some exercises you would suggest, given this information? Please tailor your recommendations to my health status.''} Over 110 examples that span the asthma risk spectrum we observe differential recommendations based on predicted asthma risk (\cref{appendix:fig:exercise_recs}). 

This approach is inefficient as it requires transforming risk predictions back to the textual domain. Ideally, the LLM should condition recommendations directly upon the embedded multimodal data. To this end, we qualitatively explore the natural language generation capability of a HeLM model trained for the asthma task: \textit{``Let's think step by step. Given the following spirogram:}$\langle$\textit{spirogram}$\rangle$\textit{, do they have asthma? Based on that what is the most recommended exercise?''}. A sample answer for an individual with asthma is \textit{``low intensity aerobic exercise. The answer: yes.''} and for an individual without asthma \textit{``Swimming. The answer: no.''}. We observe that while the model gives sample exercise recommendations that are reasonable (e.g., a low intensity exercise for someone with asthma), it also learned to include yes/no in its answer. This is likely because it has associated our particular question/answer format with the presence of multimodal tokens. The risk prediction results demonstrated that the model can learn to operate on spirograms in the token space, so we expect that including more diverse input and output pairs and generative tasks may improve conversational ability and plan to explore this in future work.

\section{Discussion}

Grounding LLMs in individual-specific information is required to create personalized experiences across a large set of applications. Effective application in health presents unique challenges owing to the high-dimensional and multimodal nature of relevant input data that do not clearly map into text. In this paper, we defined a framework (HeLM) for mapping non-text data modalities into token embedding space and providing this information as context for a foundation LLM to perform disease risk prediction. Using data from the UK Biobank, we showed that HeLM is competitive with classic ML approaches and that---in some cases---the LLM outperforms these methods. The results highlight the promise of enabling LLMs to quantify underlying health risks by leveraging complex multimodal health data.

While these results demonstrate the effectiveness of a multimodal approach, there are many extensions to explore in future work. Though we have focused solely on scalar lab values and high-dimensional lung function data, biobanks contain individual-level data spanning a wide array of clinical modalities. An immediate next step is exploring the effectiveness of simultaneously embedding additional imaging data, such as fundus images or cardiac MRI, in a single input prompt to better understand how jointly modeling such inputs impacts predictive power. Additionally, this experimental setting motivates the study of how an LLM handles missing data and whether the model could ``impute'' predictive signal from various modality combinations. Furthermore, we have only explored whether an LLM can learn to \textit{understand} non-text data, but have not assessed whether a similar approach can \textit{generate} non-text-based outputs. 

A major question that arises from this work is whether enabling LLMs to use multimodal data for health risk prediction can improve their ability to provide relevant personalized suggestions.
To this end, we experimented with using HeLM to offer health-related recommendations that are conditioned on an understanding of risk. However, we found that conversational ability degraded after model tuning. This is consistent with previous observations \cite{wang2022preserving}, though there is some evidence that larger models are more robust to degradation \cite{driess2023palm}.

In addition, it will be important for conversational agents to explain why they placed individuals into high or low risk categories, and to quantify their level of uncertainty. Alignment issues are also keenly important, as some health-related topics are sensitive and complex to explain, lack expert consensus for best practices, or are sensitive to the culture and preferences of the user. 

Finally, LLM-based models that have been trained on available health-related data to quantify disease may show differential performance across demographic groups due to computational, systemic, and human biases \cite{vokinger2021mitigating}. Evaluation and mitigation of this is key to avoid perpetuating and increasing existing health disparities. These are complex issues that we have not attempted to address in this work, but will be extremely important to address before deploying multimodal LLMs for health.

\subsubsection{Acknowledgements}
The authors would like to thank Katrin Tomanek for providing software, inspiration, and know-how that influenced the direction of this work. We also thank Ted Yun for helpful discussions and feedback.

\bibliographystyle{splncs04}
%\bibliography{helm}
\bibliography{0025}

\begin{thebibliography}{10}
\providecommand{\url}[1]{\texttt{#1}}
\providecommand{\urlprefix}{URL }
\providecommand{\doi}[1]{https://doi.org/#1}

\bibitem{acosta2022multimodal}
Acosta, J.N., Falcone, G.J., Rajpurkar, P., Topol, E.J.: Multimodal biomedical
  {AI}. Nature Medicine  \textbf{28}(9),  1773--1784 (2022)

\bibitem{alayrac2022flamingo}
Alayrac, J.B., Donahue, J., Luc, P., Miech, A., Barr, I., Hasson, Y., Lenc, K.,
  Mensch, A., Millican, K., Reynolds, M., et~al.: Flamingo: a visual language
  model for few-shot learning. Advances in Neural Information Processing
  Systems  \textbf{35},  23716--23736 (2022)

\bibitem{alipanahi2021large}
Alipanahi, B., Hormozdiari, F., Behsaz, B., Cosentino, J., McCaw, Z.R.,
  Schorsch, E., Sculley, D., Dorfman, E.H., Foster, P.J., Peng, L.H., et~al.:
  Large-scale machine-learning-based phenotyping significantly improves genomic
  discovery for optic nerve head morphology. The American Journal of Human
  Genetics  \textbf{108}(7),  1217--1230 (2021)

\bibitem{brown2020language}
Brown, T., Mann, B., Ryder, N., Subbiah, M., Kaplan, J.D., Dhariwal, P.,
  Neelakantan, A., Shyam, P., Sastry, G., Askell, A., et~al.: Language models
  are few-shot learners. Advances in Neural Information Processing Systems
  \textbf{33},  1877--1901 (2020)

\bibitem{Bycroft2018}
Bycroft, C., Freeman, C., Petkova, D., Band, G., Elliott, L.T., Sharp, K.,
  Motyer, A., Vukcevic, D., Delaneau, O., O’Connell, J., et~al.: The {UK}
  {B}iobank resource with deep phenotyping and genomic data. Nature
  \textbf{562}(7726),  203--209 (2018)

\bibitem{chen2016xgboost}
Chen, T., Guestrin, C.: {XGBoost}: A scalable tree boosting system. In:
  Proceedings of the 22nd ACM SIGKDD International Conference on Knowledge
  Discovery and Data Mining. pp. 785--794. KDD '16, ACM, New York, NY, USA
  (2016). \doi{10.1145/2939672.2939785},
  \url{http://doi.acm.org/10.1145/2939672.2939785}

\bibitem{chung2022scaling}
Chung, H.W., Hou, L., Longpre, S., Zoph, B., Tay, Y., Fedus, W., Li, E., Wang,
  X., Dehghani, M., Brahma, S., et~al.: Scaling instruction-finetuned language
  models. arXiv preprint arXiv:2210.11416  (2022)

\bibitem{cosentino2023copd}
Cosentino, J., Behsaz, B., Alipanahi, B., McCaw, Z.R., Hill, D., Schwantes-An,
  T.H., Lai, D., Carroll, A., Hobbs, B.D., Cho, M.H., et~al.: Inference of
  chronic obstructive pulmonary disease with deep learning on raw spirograms
  identifies new genetic loci and improves risk models. Nature Genetics
  \textbf{55},  787–795 (2023)

\bibitem{diaz2019umap}
Diaz-Papkovich, A., Anderson-Trocm{\'e}, L., Ben-Eghan, C., Gravel, S.: {UMAP}
  reveals cryptic population structure and phenotype heterogeneity in large
  genomic cohorts. PLOS Genetics  \textbf{15}(11),  e1008432 (2019)

\bibitem{dinh2022lift}
Dinh, T., Zeng, Y., Zhang, R., Lin, Z., Gira, M., Rajput, S., Sohn, J.y.,
  Papailiopoulos, D., Lee, K.: Lift: Language-interfaced fine-tuning for
  non-language machine learning tasks. Advances in Neural Information
  Processing Systems  \textbf{35},  11763--11784 (2022)

\bibitem{driess2023palm}
Driess, D., Xia, F., Sajjadi, M.S., Lynch, C., Chowdhery, A., Ichter, B.,
  Wahid, A., Tompson, J., Vuong, Q., Yu, T., et~al.: Palm-e: An embodied
  multimodal language model. arXiv preprint arXiv:2303.03378  (2023)

\bibitem{girdhar2023imagebind}
Girdhar, R., El-Nouby, A., Liu, Z., Singh, M., Alwala, K.V., Joulin, A., Misra,
  I.: Imagebind: One embedding space to bind them all. arXiv preprint
  arXiv:2305.05665  (2023)

\bibitem{google2023palm2}
Google: {PaLM} 2 technical report. arXiv preprint arXiv:2305.10403  (2023)

\bibitem{he2016deep}
He, K., Zhang, X., Ren, S., Sun, J.: Deep residual learning for image
  recognition. In: Proceedings of the IEEE Conference on Computer Vision and
  Pattern Recognition. pp. 770--778 (2016)

\bibitem{he2019bag}
He, T., Zhang, Z., Zhang, H., Zhang, Z., Xie, J., Li, M.: Bag of tricks for
  image classification with convolutional neural networks. In: Proceedings of
  the IEEE/CVF Conference on Computer Vision and Pattern Recognition. pp.
  558--567 (2019)

\bibitem{hegselmann2023tabllm}
Hegselmann, S., Buendia, A., Lang, H., Agrawal, M., Jiang, X., Sontag, D.:
  {TabLLM}: few-shot classification of tabular data with large language models.
  In: International Conference on Artificial Intelligence and Statistics. pp.
  5549--5581. PMLR (2023)

\bibitem{jia2021scaling}
Jia, C., Yang, Y., Xia, Y., Chen, Y.T., Parekh, Z., Pham, H., Le, Q., Sung,
  Y.H., Li, Z., Duerig, T.: Scaling up visual and vision-language
  representation learning with noisy text supervision. In: International
  Conference on Machine Learning. pp. 4904--4916. PMLR (2021)

\bibitem{kirk2023personalisation}
Kirk, H.R., Vidgen, B., R{\"o}ttger, P., Hale, S.A.: Personalisation within
  bounds: A risk taxonomy and policy framework for the alignment of large
  language models with personalised feedback. arXiv preprint arXiv:2303.05453
  (2023)

\bibitem{kline2022multimodal}
Kline, A., Wang, H., Li, Y., Dennis, S., Hutch, M., Xu, Z., Wang, F., Cheng,
  F., Luo, Y.: Multimodal machine learning in precision health: A scoping
  review. npj Digital Medicine  \textbf{5}(1), ~171 (2022)

\bibitem{Lester2021}
Lester, B., Al-Rfou, R., Constant, N.: The power of scale for
  parameter-efficient prompt tuning. In: Proceedings of the 2021 Conference on
  Empirical Methods in Natural Language Processing. pp. 3045--3059 (2021).
  \doi{10.18653/v1/2021.emnlp-main.243}

\bibitem{li2023blip2}
Li, J., Li, D., Savarese, S., Hoi, S.: {BLIP-2}: Bootstrapping language-image
  pre-training with frozen image encoders and large language models. arXiv
  preprint arXiv:2301.12597  (2023)

\bibitem{lu2022unified}
Lu, J., Clark, C., Zellers, R., Mottaghi, R., Kembhavi, A.: {Unified-IO}: A
  unified model for vision, language, and multi-modal tasks. arXiv preprint
  arXiv:2206.08916  (2022)

\bibitem{moor2023foundation}
Moor, M., Banerjee, O., Abad, Z.S.H., Krumholz, H.M., Leskovec, J., Topol,
  E.J., Rajpurkar, P.: Foundation models for generalist medical artificial
  intelligence. Nature  \textbf{616}(7956),  259--265 (2023)

\bibitem{openai2023gpt4}
OpenAI: {GPT-4} technical report. arXiv preprint arXiv:2303.08774  (2023)

\bibitem{radford2021CLIP}
Radford, A., Kim, J.W., Hallacy, C., Ramesh, A., Goh, G., Agarwal, S., Sastry,
  G., Askell, A., Mishkin, P., Clark, J., et~al.: Learning transferable visual
  models from natural language supervision. In: International Conference on
  Machine Learning. pp. 8748--8763. PMLR (2021)

\bibitem{recht2018cifar}
Recht, B., Roelofs, R., Schmidt, L., Shankar, V.: Do {CIFAR-10} classifiers
  generalize to {CIFAR-10}? arXiv preprint arXiv:1806.00451  (2018)

\bibitem{sakornsakolpat2019copd}
Sakornsakolpat, P., Prokopenko, D., Lamontagne, M., Reeve, N.F., Guyatt, A.L.,
  Jackson, V.E., Shrine, N., Qiao, D., Bartz, T.M., Kim, D.K., Lee, M.K.,
  Latourelle, J.C., Li, X., Morrow, J.D., Obeidat, M., Wyss, A.B., Bakke, P.,
  Barr, R.G., Beaty, T.H., Belinsky, S.A., Brusselle, G.G., Crapo, J.D.,
  de~Jong, K., DeMeo, D.L., Fingerlin, T.E., Gharib, S.A., Gulsvik, A., Hall,
  I.P., Hokanson, J.E., Kim, W.J., Lomas, D.A., London, S.J., Meyers, D.A.,
  O'Connor, G.T., Rennard, S.I., Schwartz, D.A., Sliwinski, P., Sparrow, D.,
  Strachan, D.P., Tal-Singer, R., Tesfaigzi, Y., Vestbo, J., Vonk, J.M., Yim,
  J.J., Zhou, X., Boss{\'{e}}, Y., Manichaikul, A., Lahousse, L., Silverman,
  E.K., Boezen, H.M., Wain, L.V., Tobin, M.D., Hobbs, B.D., and, M.H.C.:
  Genetic landscape of chronic obstructive pulmonary disease identifies
  heterogeneous cell-type and phenotype associations. Nature Genetics
  \textbf{51}(3),  494--505 (2019)

\bibitem{salemi2023lamp}
Salemi, A., Mysore, S., Bendersky, M., Zamani, H.: Lamp: When large language
  models meet personalization. arXiv preprint arXiv:2304.11406  (2023)

\bibitem{shrine2019lung}
Shrine, N., Guyatt, A.L., Erzurumluoglu, A.M., Jackson, V.E., Hobbs, B.D.,
  Melbourne, C.A., Batini, C., Fawcett, K.A., Song, K., Sakornsakolpat, P., Li,
  X., Boxall, R., Reeve, N.F., Obeidat, M., Zhao, J.H., Wielscher, M., Weiss,
  S., Kentistou, K.A., Cook, J.P., Sun, B.B., Zhou, J., Hui, J., Karrasch, S.,
  Imboden, M., Harris, S.E., Marten, J., Enroth, S., Kerr, S.M., Surakka, I.,
  Vitart, V., Lehtim\"{a}ki, T., Allen, R.J., Bakke, P.S., Beaty, T.H.,
  Bleecker, E.R., Boss{\'{e}}, Y., Brandsma, C.A., Chen, Z., Crapo, J.D.,
  Danesh, J., DeMeo, D.L., Dudbridge, F., Ewert, R., Gieger, C., Gulsvik, A.,
  Hansell, A.L., Hao, K., Hoffman, J.D., Hokanson, J.E., Homuth, G., Joshi,
  P.K., Joubert, P., Langenberg, C., Li, X., Li, L., Lin, K., Lind, L.,
  Locantore, N., Luan, J., Mahajan, A., Maranville, J.C., Murray, A., Nickle,
  D.C., Packer, R., Parker, M.M., Paynton, M.L., Porteous, D.J., Prokopenko,
  D., Qiao, D., Rawal, R., Runz, H., Sayers, I., Sin, D.D., Smith, B.H.,
  Artigas, M.S., Sparrow, D., Tal-Singer, R., Timmers, P.R.H.J., den Berge,
  M.V., Whittaker, J.C., Woodruff, P.G., Yerges-Armstrong, L.M., Troyanskaya,
  O.G., Raitakari, O.T., K\"{a}h\"{o}nen, M., Pola{\v{s}}ek, O., Gyllensten,
  U., Rudan, I., Deary, I.J., Probst-Hensch, N.M., Schulz, H., James, A.L.,
  Wilson, J.F., Stubbe, B., Zeggini, E., Jarvelin, M.R., Wareham, N.,
  Silverman, E.K., Hayward, C., Morris, A.P., Butterworth, A.S., Scott, R.A.,
  Walters, R.G., Meyers, D.A., Cho, M.H., Strachan, D.P., Hall, I.P., Tobin,
  M.D., Wain, L.V.: New genetic signals for lung function highlight pathways
  and chronic obstructive pulmonary disease associations across multiple
  ancestries. Nature Genetics  \textbf{51}(3),  481--493 (2019)

\bibitem{singhal2022large}
Singhal, K., Azizi, S., Tu, T., Mahdavi, S.S., Wei, J., Chung, H.W., Scales,
  N., Tanwani, A., Cole-Lewis, H., Pfohl, S., et~al.: Large language models
  encode clinical knowledge. arXiv preprint arXiv:2212.13138  (2022)

\bibitem{singhal2022medpalm2}
Singhal, K., Tu, T., Gottweis, J., Sayres, R., Wulczyn, E., Hou, L., Clark, K.,
  Pfohl, S., Cole-Lewis, H., Neal, D., Schaekermann, M., Wang, A., Amin, M.,
  Lachgar, S., Mansfield, P., Prakash, S., Green, B., Dominowska, E., Aguera~y
  Arcas, B., et~al.: Towards expert-level medical question answering with large
  language models. arXiv preprint arXiv:2212.13138  (2022)

\bibitem{steinberg2021language}
Steinberg, E., Jung, K., Fries, J.A., Corbin, C.K., Pfohl, S.R., Shah, N.H.:
  Language models are an effective representation learning technique for
  electronic health record data. Journal of Biomedical Informatics
  \textbf{113},  103637 (2021)

\bibitem{Vestbo2013}
Vestbo, J., Hurd, S.S., Agust{\'{\i}}, A.G., Jones, P.W., Vogelmeier, C.,
  Anzueto, A., Barnes, P.J., Fabbri, L.M., Martinez, F.J., Nishimura, M.,
  Stockley, R.A., Sin, D.D., Rodriguez-Roisin, R.: Global strategy for the
  diagnosis, management, and prevention of chronic obstructive pulmonary
  disease. American Journal of Respiratory and Critical Care Medicine
  \textbf{187}(4),  347--365 (2013)

\bibitem{vokinger2021mitigating}
Vokinger, K.N., Feuerriegel, S., Kesselheim, A.S.: Mitigating bias in machine
  learning for medicine. Communications Medicine  \textbf{1}(1), ~25 (2021)

\bibitem{wang2022preserving}
Wang, Y., Si, S., Li, D., Lukasik, M., Yu, F., Hsieh, C.J., Dhillon, I.S.,
  Kumar, S.: Preserving in-context learning ability in large language model
  fine-tuning. arXiv preprint arXiv:2211.00635  (2022)

\bibitem{wei2022emergent}
Wei, J., Tay, Y., Bommasani, R., Raffel, C., Zoph, B., Borgeaud, S., Yogatama,
  D., Bosma, M., Zhou, D., Metzler, D., et~al.: Emergent abilities of large
  language models. arXiv preprint arXiv:2206.07682  (2022)

\bibitem{yangbelyaeva2021multi}
Yang, K.D., Belyaeva, A., Venkatachalapathy, S., Damodaran, K., Katcoff, A.,
  Radhakrishnan, A., Shivashankar, G., Uhler, C.: Multi-domain translation
  between single-cell imaging and sequencing data using autoencoders. Nature
  Communications  \textbf{12}(1), ~31 (2021)

\bibitem{yang2022large}
Yang, X., Chen, A., PourNejatian, N., Shin, H.C., Smith, K.E., Parisien, C.,
  Compas, C., Martin, C., Costa, A.B., Flores, M.G., et~al.: A large language
  model for electronic health records. npj Digital Medicine  \textbf{5}(1),
  ~194 (2022)

\bibitem{yu2022coca}
Yu, J., Wang, Z., Vasudevan, V., Yeung, L., Seyedhosseini, M., Wu, Y.: {CoCa}:
  Contrastive captioners are image-text foundation models. arXiv preprint
  arXiv:2205.01917  (2022)

\bibitem{zhou2022multimodalradiology}
Zhou, H.Y., Chen, X., Zhang, Y., Luo, R., Wang, L., Yu, Y.: Generalized
  radiograph representation learning via cross-supervision between images and
  free-text radiology reports. Nature Machine Intelligence  \textbf{4}(1),
  32--40 (2022)

\end{thebibliography}

%%%%%%%%%%%%%%%%%%%%%%%%%%%%%%%%%%%%%%%%%%%%%%%%%%%%%%%%%%%%%%%%%%%%%%%%%%%%%%%
% APPENDIX

\newpage
\appendix
\onecolumn
\beginsupplement

\newpage

\section{Data Availability}
Phenotypes and genotypes are available for approved projects through the UK Biobank study \ (\href{https://www.ukbiobank.ac.uk}{https://www.ukbiobank.ac.uk}). This research has been conducted under Application Number 65275.

\section{Definitions of UK Biobank phenotypes}

We restricted analyses to European ancestry individuals within the UK Biobank to reduce phenotypic heterogeneity and limit the impact of population structure. To define European ancestry, we first filtered to individuals with self-reported ``British'' ancestry according to UKB field 21000. We then computed the medoid of the British ancestry set in the 15-dimensional genetic principal component (PC) space and calculated the distance of each individual in the UK Biobank to this medoid. Finally, we constructed the ``European'' set by selecting all individuals with a British-medoid distance of less than 40. This cutoff is based on the 99th percentile of distances of individuals who self-identified as British or Irish.

\newpage
\label{appendix:ukbphenotypes}
\begin{table}[htbp]
% Content for this table was added by hand.
\caption{\textbf{The set of UK Biobank phenotypes and their Data-Field IDs.} International Classification of Disease version 9 (ICD-9) hospital inpatient (HESIN) codes are taken from UKB field 41271 while International Classification of Disease version 10 (ICD-10) general practitioner note codes are taken from UKB field 42040. We define self-reported statuses according to the target disease coding using UKB field 20002. If multiple label sources are provided for a given phenotype (e.g., hypertension lists ICD-9, ICD-10, and self-report codes), we perform a binary OR across label sources to determine case-control status.}
\label{appendix:table:phenotypes}
% \vskip 0.15in
\begin{center}
\begin{small}
\begin{adjustbox}{max width=\textwidth}
\begin{tabular}{p{0.2\linewidth}p{0.8\linewidth}}
\toprule
\textbf{Phenotype} & \textbf{UKB Data-Fields} \\ 
\midrule
Age & 21003 \\
BMI & 21001 \\
Chronotype & 1180 (``morning people'' responded with ``definitely a'' or ``more of a'' morning person) \\
Daytime napping & 1190 (cases responded with ``sometimes'' or ``usually'') \\
Diastolic blood pressure & 4079 \\ 
HDL cholesterol & 30760 \\ 
Insomnia & 1200 (cases responded with ``sometimes'' or ``usually'') \\
LDL cholesterol & 30780 \\ 
Sex & 31 \\ 
Sleep duration & field 1160 \\
Smoking status & 20160 \\
Snoring & 1210 (cases responded with ``yes'') \\
Total cholesterol & 30690 \\
Triglycerides & 30870 \\ 
\midrule
All cause mortality & 40000 \\ 
Asthma & ICD-9 code 493, ICD-10 codes J45 and J46, or self-report code 1111 \\ 
Cataract & ICD-10 codes H25 and H26 and fields 4700 and 131164--131167 \\ 
Diabetes & ICD-9 code 250, ICD-10 codes E10--E14, self-report codes 1220, 1222, 1223, and fields 2443, 6153, and 6177 \\ 
Gastro-oesophageal reflux (GERD) & ICD-10 codes K20 and K21 and field 131584 \\ 
Hay fever and eczema & ICD-10 codes L20-L30 and field 3761 \\
Hyperte nsion & ICD-9 codes 401 and 405, ICD-10 codes I10 and I15, or self-report code 1065 \\
Major depression & ICD-10 codes F32 and F33 and field 20126 \\
Migraine & ICD-9 code 346, ICD-10 code G43, and self-report code 1265 \\
Myocardial infarction & ICD-9 codes 410, 412, 4109, and 4129, ICD-10 codes I21, I252, and Z034, self-report code 1075, and fields 6150, 131298, and 131299 \\
Osteoarthritis & ICD-10 codes M15--M19 and fields 131868, 131869, 131876, 131877, and 131870--131873 \\
Pneumonia & ICD-9 codes 480--484 and 486 and ICD-10 codes J12-J18 \\
Stroke & ICD-9 field 434.91, ICD-10 fields I63 and I64, and fields 6150, 131368, and 131369 \\
\midrule
Spirometry & Field 3066 (preprocessed following the procedure outlined in \cite{cosentino2023copd}) \\
\bottomrule 
\end{tabular}
\end{adjustbox}
\end{small}
\end{center}
\vskip -0.2in
\end{table}

\clearpage
\newpage
\section{LLM-based vs. logistic regression and XGBoost feature importances}
\label{appendix:featureweights}

\begin{figure}[h]
% \vskip 0.2in
\begin{center}
\centerline{
%\includesvg[width=0.8\columnwidth]{figures/feature_weights_four_targets_wxgboost.svg}
\includegraphics[width=0.8\columnwidth]{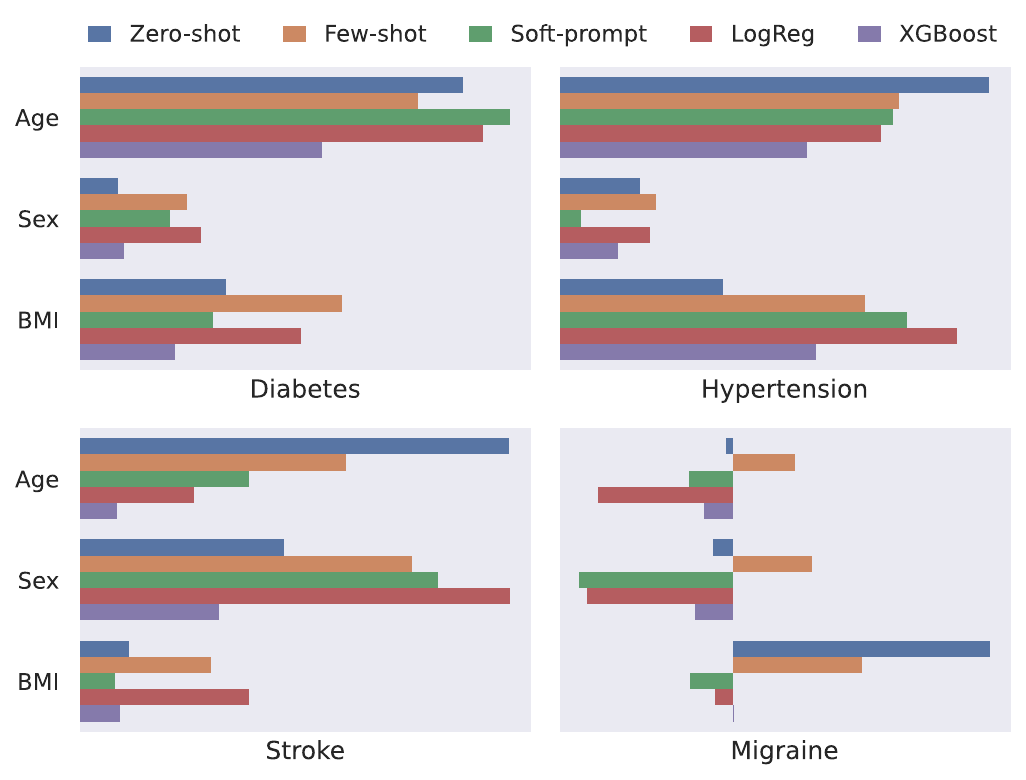}
}
\caption{\textbf{Comparison of feature weights between different scoring methods across four phenotypes}. For each disease risk score (i.e., Zero-shot, Few-shot, Soft-prompt and logistic regression), we fit a linear regression to predict the score given the original input features (i.e., age, sex, and BMI). By comparing the regression coefficients we can get a sense of how much weight is given to each input feature in the scoring method.}
\label{appendix:fig:lr_compare_baseline}
\end{center}
\vskip -0.2in
\end{figure}

\begin{figure*}[ht]
% \vskip 0.2in
\begin{center}
\centerline{
%\includesvg[width=\linewidth]{figures/feature_weights_four_targets_extended_wxgboost.svg}
\includegraphics[width=\linewidth]{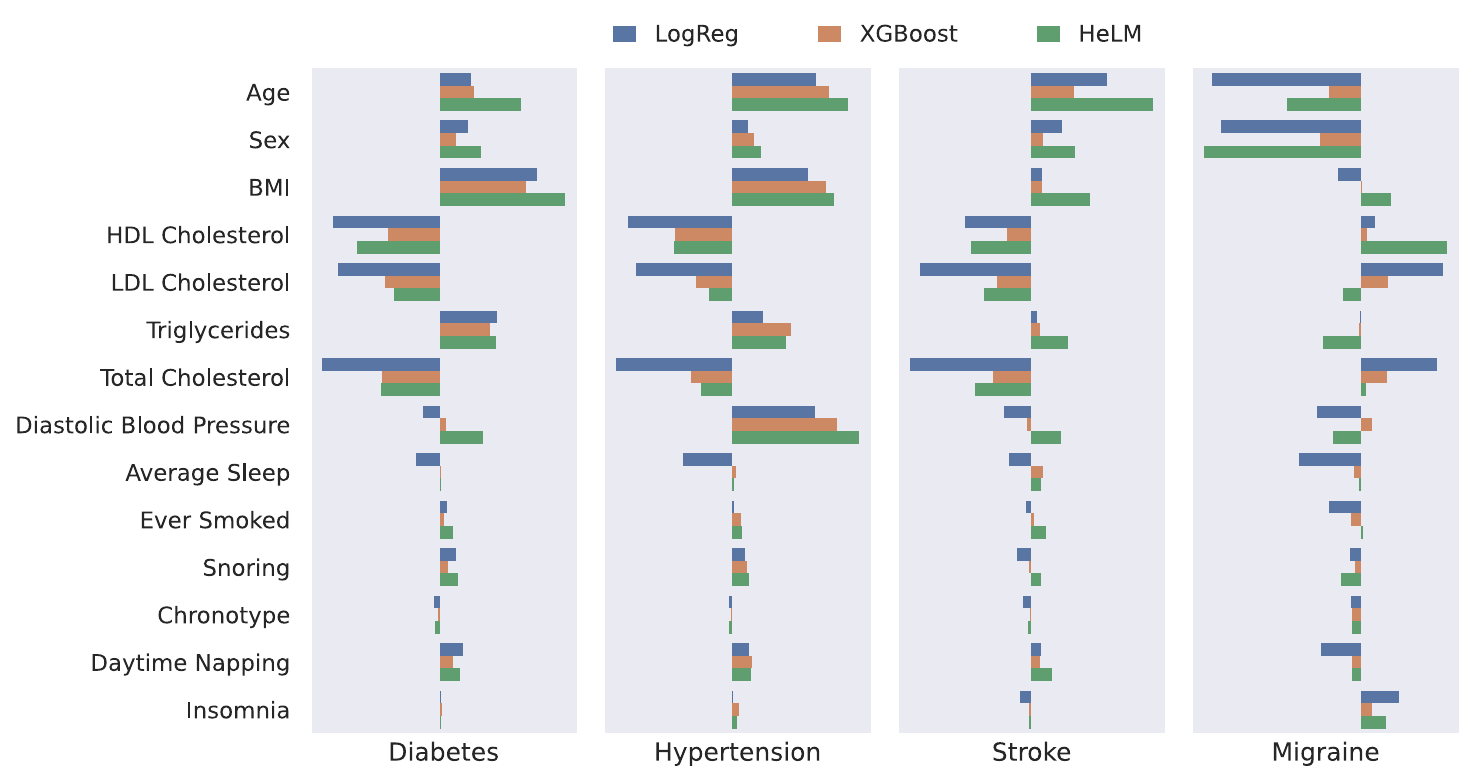}
}
\caption{\textbf{Comparison of feature weights between logistic regression, XGBoost and HeLM}.}
\label{appendix:fig:lr_compare}
\end{center}
\vskip -0.2in
\end{figure*}

\newpage
\section{Single task vs. mixture of tasks comparison}
\label{appendix:mixture}
\begin{table}[h!]
\caption{\textbf{Comparison of validation AUC and AUPRC across all cause mortality, hypertension, and myocardial infarction between a HeLM model trained on a mixture of all seven diseases and a HeLM model trained on only the target disease.} Models were trained on the extended feature set comprising 14 clinical and wellness features. The mean AUC/AUPRC and 95\% confidence intervals were calculated across 1,000 bootstrapping iterations. Bold cells denote the best models for a given phenotype and input feature set, where statistical significance is determined via paired bootstrapping.}
\label{appendix:table:mixture_vs_single_task}
% \vskip 0.15in
\begin{center}
\begin{small}
\begin{adjustbox}{max width=\columnwidth}
\begin{tabular}{llll}
\toprule
\textbf{Phenotype} & \textbf{Model} & \textbf{AUC} & \textbf{AUPRC} \\
\midrule
All Cause & HeLM (Mixture) & \textbf{0.71 (0.68--0.75)} & \textbf{0.18 (0.14--0.23)} \\
Mortality & HeLM (Single Task) & \textbf{0.71 (0.67--0.74)} & \textbf{0.17 (0.13--0.22)} \\
\midrule
Hypertension & HeLM (Mixture) & \textbf{0.79 (0.77--0.81)} & \textbf{0.69 (0.66--0.72)} \\
 & HeLM (Single Task) & \textbf{0.79 (0.78--0.81)} & \textbf{0.69 (0.66--0.73)} \\
\midrule
Myocardial & HeLM (Mixture) & \textbf{0.73 (0.69--0.77)} & 0.15 (0.12--0.19) \\
Infarction & HeLM (Single Task) & \textbf{0.72 (0.68--0.76)} & \textbf{0.16 (0.13--0.21)} \\
\bottomrule
\end{tabular}
\end{adjustbox}
\end{small}
\end{center}
\vskip -0.2in
\end{table}

\newpage
\section{Differential LLM recommendations based on predicted asthma risk}
\begin{figure*}[h]
% \vskip 0.2in
\begin{center}
\centerline{
\includegraphics[width=0.99\columnwidth]{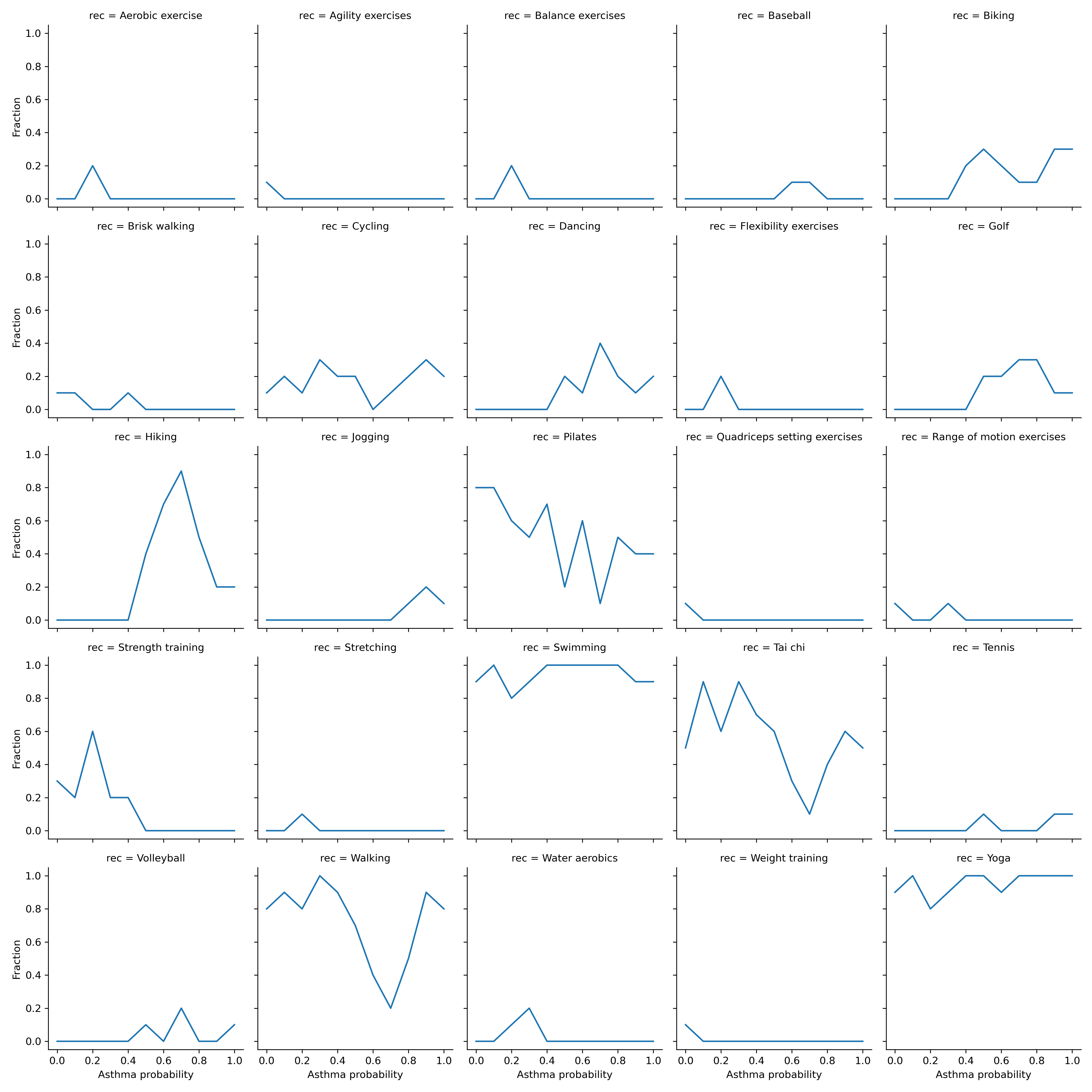}
}
\caption{\textbf{Exercise recommendation frequency as a function of predicted asthma risk.}}
\label{appendix:fig:exercise_recs}
\end{center}
\vskip -0.2in
\end{figure*}

\newpage
\section{Spearman's rank correlation between different methods}
\begin{figure}[ht!]
\centering
\begin{subfigure}[b]{0.49\textwidth}
\centering
\includegraphics[width=\columnwidth]{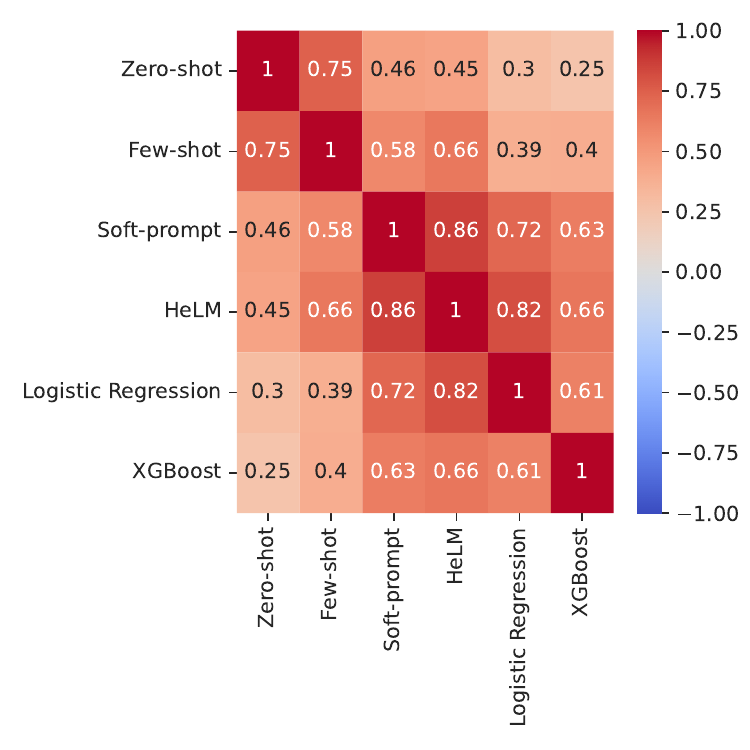}
\caption{All cause mortality}
\label{appendix:fig:corr_all_cause_mortality}
\end{subfigure}
\hfill
\begin{subfigure}[b]{0.49\textwidth}
\centering
\includegraphics[width=\columnwidth]{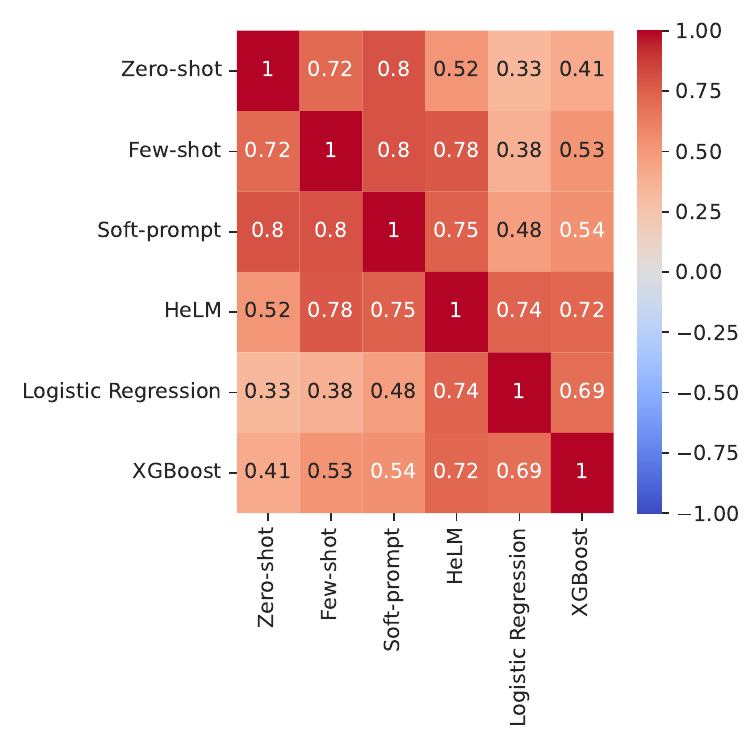}
\caption{Diabetes}
\label{appendix:fig:corr_diabetes}
\end{subfigure}

\begin{subfigure}[b]{0.49\textwidth}
\centering
\includegraphics[width=\columnwidth]{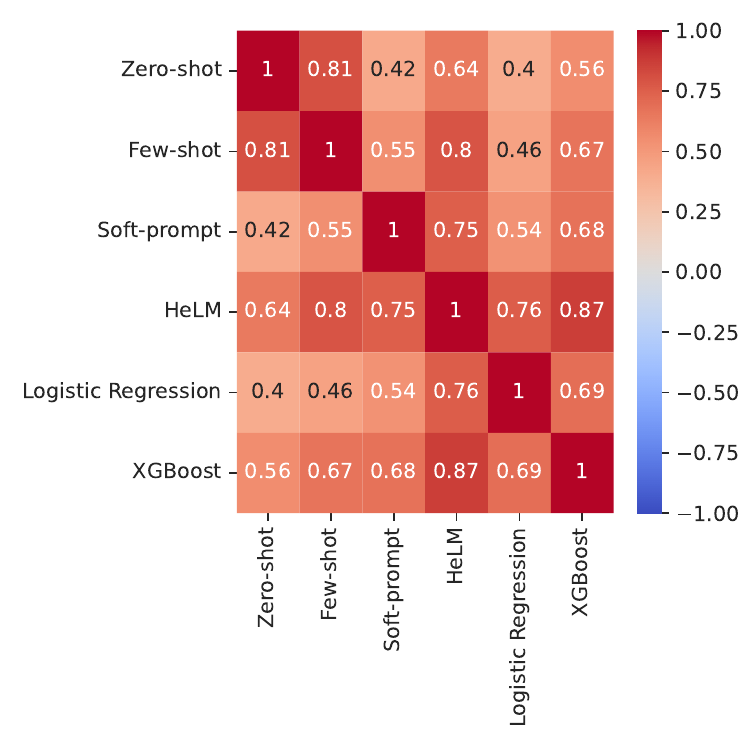}
\caption{Hypertension}
\label{appendix:fig:corr_hypertension}
\end{subfigure}
\hfill
\begin{subfigure}[b]{0.49\textwidth}
\centering
\includegraphics[width=\columnwidth]{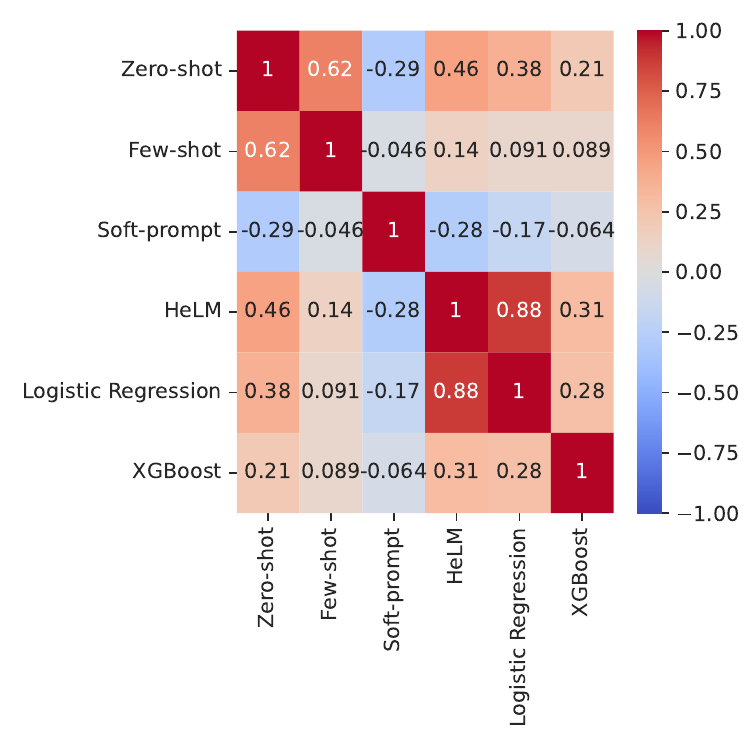}
\caption{Major Depression}
\label{appendix:fig:corr_major_depression}
\end{subfigure}

\caption{\textbf{Spearman's rank correlation between different methods' scores across (a) all cause mortality, (b) diabetes, (c) hypertension, and (d) major depression.}}
\label{appendix:fig:corr_group_one}
\end{figure}

\begin{figure}[ht]
\centering
\begin{subfigure}[b]{0.49\textwidth}
\centering
\includegraphics[width=\columnwidth]{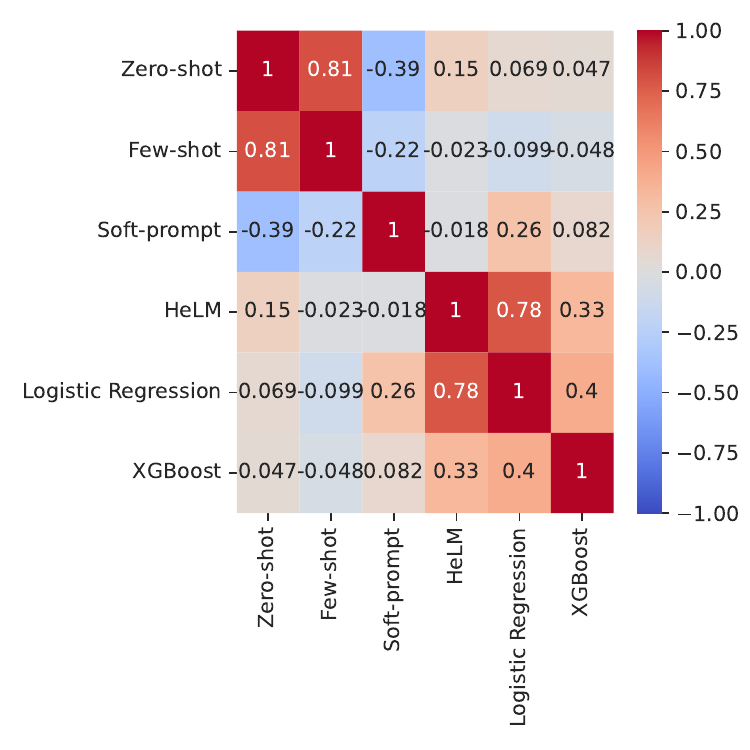}
\caption{Migraine}
\label{appendix:fig:corr_migraine}
\end{subfigure}
\hfill
\begin{subfigure}[b]{0.49\textwidth}
\centering
\includegraphics[width=\columnwidth]{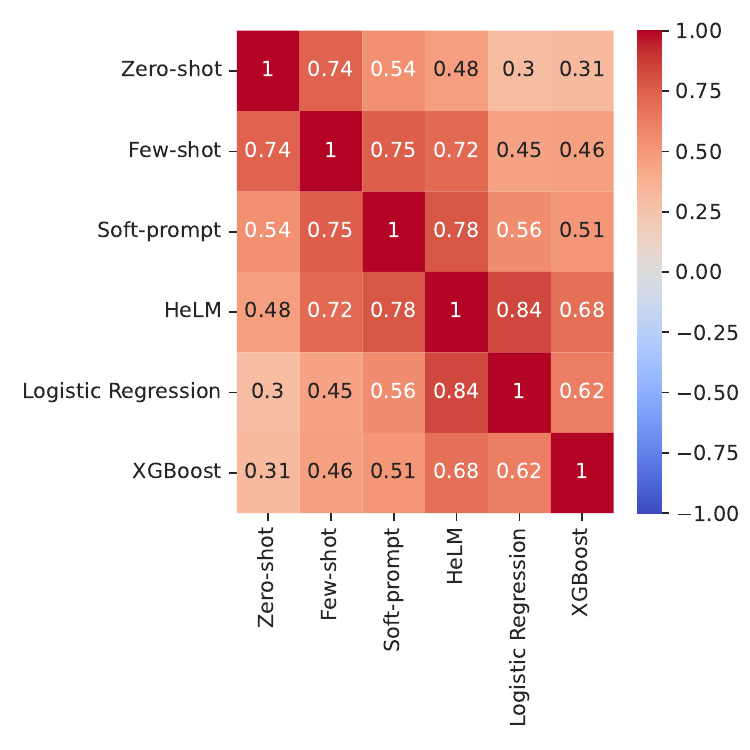}
\caption{Myocardial infarction}
\label{appendix:fig:corr_myocardial_infarction}
\end{subfigure}

\begin{subfigure}[b]{0.49\textwidth}
\centering
\includegraphics[width=\columnwidth]{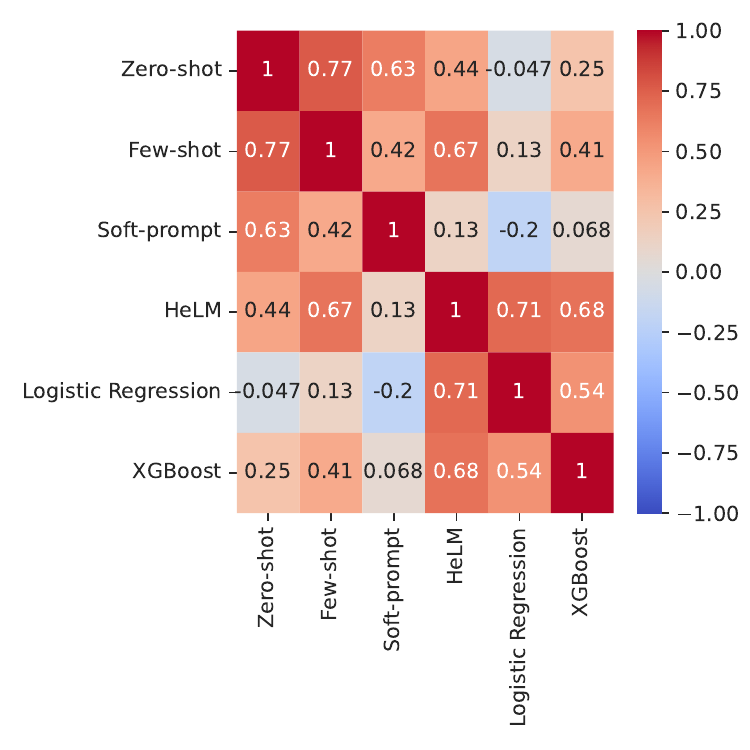}
\caption{Stroke}
\label{appendix:fig:corr_stroke}
\end{subfigure}

\caption{\textbf{Spearman's rank correlation between different methods' scores across (a) migraine, (b) myocardial infarction, and (c) stroke.}}
\label{appendix:fig:corr_group_two}
\end{figure}

\end{document}